\documentclass[11pt]{article}

\usepackage{epsfig,multicol,amsmath}

\RequirePackage{xspace}

\usepackage{relsize}
\def\babar{\mbox{\slshape B\kern-0.1em{\smaller A}\kern-0.1em
    B\kern-0.1em{\smaller A\kern-0.2em R}}}

\usepackage{relsize}
\def\belle{\mbox{B\kern-0.05em E\kern-0.05em L\kern-0.05em L\kern-0.05em E}}

\usepackage{relsize}
\def\cleo{\mbox{C\kern-0.05em L\kern-0.05em E\kern-0.05em O}}

\def\epem       {\ensuremath{e^+e^-}\xspace}

\def\qqbar {\ensuremath{q\overline q}\xspace}
\def\u     {\ensuremath{u}\xspace}
\def\ubar  {\ensuremath{\overline u}\xspace}

\def\d     {\ensuremath{d}\xspace}
\def\dbar  {\ensuremath{\overline d}\xspace}

\def\s     {\ensuremath{s}\xspace}
\def\sbar  {\ensuremath{\overline s}\xspace}

\def\c     {\ensuremath{c}\xspace}
\def\cbar  {\ensuremath{\overline c}\xspace}

\def\b     {\ensuremath{b}\xspace}
\def\bbar  {\ensuremath{\overline b}\xspace}

\def\Kbar  {\kern 0.2em\overline{\kern -0.2em K}{}\xspace}

\def\Kz    {\ensuremath{K^0}\xspace}
\def\Kzb   {\ensuremath{\Kbar^0}\xspace}
\def\KzKzb {\ensuremath{\Kz \kern -0.16em \Kzb}\xspace}
\def\Kp    {\ensuremath{K^+}\xspace}
\def\Km    {\ensuremath{K^-}\xspace}

\def\KpKm  {\ensuremath{\Kp \kern -0.16em \Km}\xspace}
\def\KS    {\ensuremath{K^0_{\scriptscriptstyle S}}\xspace}
\def\KL    {\ensuremath{K^0_{\scriptscriptstyle L}}\xspace}

\def\Dbar    {\kern 0.2em\overline{\kern -0.2em D}{}\xspace}

\def\Dz      {\ensuremath{D^0}\xspace}
\def\Dzb     {\ensuremath{\Dbar^0}\xspace}
\def\DzDzb   {\ensuremath{\Dz {\kern -0.16em \Dzb}}\xspace}
\def\Dp      {\ensuremath{D^+}\xspace}
\def\Dm      {\ensuremath{D^-}\xspace}

\def\DpDm    {\ensuremath{\Dp {\kern -0.16em \Dm}}\xspace}

\def\B       {\ensuremath{B}\xspace}
\def\Bbar    {\kern 0.18em\overline{\kern -0.18em B}{}\xspace}

\def\BB      {\ensuremath{B\Bbar}\xspace}
\def\Bz      {\ensuremath{B^0}\xspace}
\def\Bzb     {\ensuremath{\Bbar^0}\xspace}
\def\BzBzb   {\ensuremath{\Bz {\kern -0.16em \Bzb}}\xspace}
\def\Bu      {\ensuremath{B^+}\xspace}
\def\Bub     {\ensuremath{B^-}\xspace}

\def\BpBm    {\ensuremath{\Bu {\kern -0.16em \Bub}}\xspace}

\def\jpsi     {\ensuremath{{J\mskip -3mu/\mskip -2mu\psi\mskip 2mu}}\xspace}
\def\psitwos  {\ensuremath{\psi{(2S)}}\xspace}

\mathchardef\Upsilon="7107
\def\Y#1S{\ensuremath{\Upsilon{(#1S)}}\xspace}

\def\FourS {\Y4S}

\newcommand{\tev}{\ensuremath{\mathrm{\,Te\kern -0.1em V}}\xspace}
\newcommand{\gev}{\ensuremath{\mathrm{\,Ge\kern -0.1em V}}\xspace}
\newcommand{\mev}{\ensuremath{\mathrm{\,Me\kern -0.1em V}}\xspace}
\newcommand{\kev}{\ensuremath{\mathrm{\,ke\kern -0.1em V}}\xspace}
\newcommand{\ev}{\ensuremath{\mathrm{\,e\kern -0.1em V}}\xspace}
\newcommand{\gevc}{\ensuremath{{\mathrm{\,Ge\kern -0.1em V\!/}c}}\xspace}
\newcommand{\mevc}{\ensuremath{{\mathrm{\,Me\kern -0.1em V\!/}c}}\xspace}
\newcommand{\gevcc}{\ensuremath{{\mathrm{\,Ge\kern -0.1em V\!/}c^2}}\xspace}
\newcommand{\mevcc}{\ensuremath{{\mathrm{\,Me\kern -0.1em V\!/}c^2}}\xspace}

\def\cm   {\ensuremath{\rm \,cm}\xspace}

\def\mm   {\ensuremath{\rm \,mm}\xspace}

\def\mum  {\ensuremath{\,\mu\rm m}\xspace}

\def\nb         {\ensuremath{\rm \,nb}\xspace}

\def\invpb {\ensuremath{\mbox{\,pb}^{-1}}\xspace}

\def\invfb   {\ensuremath{\mbox{\,fb}^{-1}}\xspace}

\def\mus  {\ensuremath{\rm \,\mus}\xspace}

\def\ps   {\ensuremath{\rm \,ps}\xspace}

\def\mus        {\ensuremath{\,\mu{\rm s}}\xspace}    
\def\ps         {\ensuremath{{\rm \,ps}}\xspace}  

\def\to                 {\ensuremath{\rightarrow}\xspace}

\def\pepii{PEP-II}
\def\kekb{KEK-B}

\def\gsim{{~\raise.15em\hbox{$>$}\kern-.85em
          \lower.35em\hbox{$\sim$}~}\xspace}
\def\lsim{{~\raise.15em\hbox{$<$}\kern-.85em
          \lower.35em\hbox{$\sim$}~}\xspace}

\def\CP                {\ensuremath{C\!P}\xspace}
\def\T                 {\ensuremath{T}\xspace}
\def\CPT               {\ensuremath{C\!PT}\xspace} 

\def\stwob{\ensuremath{\sin\! 2 \beta   }\xspace}

\def\mistag{\ensuremath{w}\xspace}

\def\deltamd{\ensuremath{{\rm \Delta}m_d}\xspace}
\def\deltambd{\ensuremath{{\rm \Delta}m_{B_d}}\xspace}
\def\deltambs{\ensuremath{{\rm \Delta}m_{B_s}}\xspace}
\def\gammabd{\ensuremath{{\rm \Gamma}_{B_d}}\xspace}
\def\deltagbd{\ensuremath{{\rm \Delta}{\gammabd}}\xspace}
\def\AfCP{\ensuremath{      {A}_{f_{CP}} } }
\def\AbfCP{\ensuremath{ \overline{A}_{f_{CP}} } }

\def\cpipi{\ensuremath{C_{\pi\pi}}}
\def\spipi{\ensuremath{S_{\pi\pi}}}

\newcommand\prl[3] {{\it Phys.\ Rev.\ Lett.\ }{\bf #1} (#2) #3}
\newcommand\ptp[3] {{\it Prog.\ Theor.\ Phys.\ }{\bf #1} (#2) #3}
\newcommand\epjc[3] {{\it Eur.\ Phys.\ J. }{\bf C #1} (#2) #3}
\newcommand\plb[3] {{\it Phys.\ Lett.\ }{\bf B #1} (#2) #3}
\newcommand\nc[3]  {{\it Nuovo Cim.\ }{\bf #1} (#2) #3}
\newcommand\prd[3] {{\it Phys.\ Rev.\ }{\bf D #1} (#2) #3}
\newcommand\npb[3] {{\it Nucl.\ Phys.\ }{\bf B #1} (#2) #3}
\newcommand\zpc[3] {{\it Z.\ Physik }{\bf C #1} (#2) #3}

\newcommand{\hepph}[1]{ {\tt hep-ph/#1}}

\newcommand{\hepex}[1]{ {\tt hep-ex/#1} }

\begin{document}

\begin{center}
\Large \bf \CP\ violation in the \B\ meson system: recent
experimental results
\end{center}
\bigskip \bigskip

\begin{center}
\large Gautier Hamel de Monchenault\\
        CEA/DSM/DAPNIA/SPP\\ Saclay, F-91191 Gif-sur-Yvette Cedex, France
\end{center}
\bigskip \bigskip

\begin{center}
Published in Proceedings of the EPS International Conference on
High Energy Physics, Budapest, 2001 (D. Horvath, P. Levai, A.
Patkos, eds.), JHEP ({\tt http://jhep.sissa.it/}) Proceedings
Section, PrHEP-hep2001/293.
\end{center}
\bigskip \bigskip

\begin{center}
\large \bf Abstract
\end{center}
Recent experimental results on \CP\ violation in the study of \B\
meson decays are reviewed. The emphasis is put on the recent
measurements of the \CP\ parameter \stwob\ by the \babar\ and
\belle\ experiments at asymmetric \B\ factories, which establish
for the first time \CP\ violation in the \B\ meson system.

\section{Introduction}

\CP\ violation has been extensively studied in the \Kz\ system
since the discovery of the phenomenon in \KL\ decays thirty seven
years ago~\cite{ref:beneke_budapest}, while the last fifteen years
have been rich in experimental and theoretical developments in
Heavy Flavor physics~\cite{ref:ligeti_budapest}. The consistency
of the experimental results with the general scheme of charged
weak interactions and \CP\ violation in the Standard Model of
particle physics is highly non-trivial.  Now is the time to
challenge the model experimentally in its prediction of large \CP\
violation effects in the \B\ meson system.  This paper gives an
overview of the present experimental knowledge on the subject. We
start with a status report on the new generation of \B\ factory
experiments, followed by an introduction to the \BzBzb\ system and
the CKM matrix, and an overview of present experimental
constraints on the Unitarity Triangle from \CP-violating $K$ and
\CP-conserving $B$ observations. We then review in turn searches
for \CP\ violation in the \B\ system: direct \CP\ violation, \CP\
violation in \BzBzb\ mixing and, finally, \CP\ violation in the
interference between mixing and decay, with an emphasis on the
recent \stwob\ measurements by \babar\ and \belle. We conclude the
review by an overview of experimental prospects in the domain.

\section{Status of \B\ factory experiments}

The three \epem\ \B\ factories in activity are CESR  at Cornell,
\pepii\ at SLAC and \kekb\ at KEK  at which \B\ meson pairs are
produced in \epem\ annihilations in the \FourS\ resonance energy
region at $\sqrt{s} \approx 10.58\gev$.

The new-generation machines, \pepii\ and \kekb, are
energy-asymmetric: the electron and positron beams are stored at
different energies in two separate storage rings so that the
proper time difference between the two \B\ meson decays can be
deduced from the measurable distance between the two decay
vertices along the boost axis.  Both machines have started
operation in late Spring 1999 and have improved steadily their
performances. At the time of the Conference in July 2001, the two
machines had already demonstrated very high instantaneous
luminosities with tolerable backgrounds for the detectors: $4.5
\times 10^{33} \cm^{-2} \s^{-1}$ for \kekb\ (for a design
luminosity of $10^{34}$) and $3.5 \times 10^{33}$ for \pepii
(already beyond the design luminosity of $3 \times 10^{33}$). The
data samples recorded by the \babar\ and \belle\ experiments were
of comparable size: about 30\invfb\ at the \FourS\ resonance
(corresponding to about 32 million \BB\ pairs).

\section{The \BzBzb\ system}

The light $B_L$ and heavy $B_H$ mass eigenstates of the neutral
$B_d$ meson system (made of \b\ and \d\ quarks) are given by
\footnote{\CPT\ invariance is assumed throughout this paper}:
\begin{equation}\label{physicalstates}
  |B_L\rangle = p\, |\Bz\rangle + q\, |\Bzb\rangle,\ \
  |B_H\rangle = p\, |\Bz\rangle - q\, |\Bzb\rangle.
\end{equation}
where \Bz\ and \Bzb\ are the flavor eigenstates of the system,
related through \CP\ transformation according to: $\CP\,
|\Bz\rangle = e^{2i\xi_B}\, |\Bzb\rangle$ (the phase $\xi_B$ is
arbitrary, due to \b-flavor conservation by strong interactions).
The complex coefficients $p$ and $q$ are normalized
$(|p|^{2}+|q|^{2}=1)$.  The phase of $q/p$, which also depends on
phase conventions, is not an observable; only the modulus of this
quantity, $|p/q|$, has a physical significance.

The mass difference \deltambd\ and width difference \deltagbd\
between the two mass eigenstates are defined as:
\begin{equation}\label{dmd}
  \deltambd\equiv m_{B_H}-m_{B_L}, \ \ \deltagbd \equiv {\rm
  \Gamma}_{B_H} - {\rm \Gamma}_{B_L} \, .
\end{equation}
Based on model-independent considerations, the two mesons are
expected to have a negligible difference in lifetime,
$\deltagbd/\gammabd \sim 10^{-2}$\cite{ref:bigi-88}. In
particular, $ \deltagbd/\gammabd \ll x_d$ where $ x_d \equiv
\deltambd / \gammabd = 0.73 \pm 0.05 $. Neglecting \deltagbd\
versus \deltambd\, the time evolution of a state prepared
initially ({\it i.e.} at time $t=0$) in a pure \Bz\ or \Bzb\
state, respectively, can be written as follows~:
\begin{eqnarray}\label{eq:timeEvolution}
  | B^0_{\rm {phys}}(t) \rangle = e^{-i\, m t }\, e^{-{\rm \Gamma\, t/
  2}} \{\ \cos{ (\, \deltambd\, t/2\, ) } \, | \Bz \rangle + i\, (q/p)\,   \sin{(\, \deltambd\,  t/2
  \,)}\,
  | \Bzb \rangle \ \}  \nonumber \\
  | \Bzb_{\rm {phys}}(t) \rangle = e^{-i\, m t }\, e^{-{\rm \Gamma\, t/
  2}} \{\ \cos{ (\, \deltambd\, t/2\, ) } \, | \Bzb \rangle + i\, (p/q
  )\,   \sin{(\, \deltambd\,  t/2
  \,)}\,
  | \Bz \rangle \ \}
\end{eqnarray}
where $m=\frac{1}{2}(m_{B_H}+m_{B_L})$ and ${\rm \Gamma}
=\frac{1}{2}({\rm \Gamma}_{B_H}+{\rm \Gamma}_{B_L})$.

\section{\CP\ violation in the Standard Model}

In the Standard Model of strong and electro-weak interactions,
\CP\ violation arises from the presence of a single irremovable
phase in the unitary complex mixing matrix for the three quark
generations~\cite{ref:ckm}\cite{ref:babar_physics_book}.  This
phase is called the Kobayashi-Maskawa phase, and the matrix, the
Cabibbo-Kobayashi-Maskawa (CKM) mixing matrix.

The unitarity of the CKM matrix can be expressed in geometric form
as six triangles of equal areas in the complex plane.  A non-zero
area directly implies the existence of a \CP-violating
phase~\cite{ref:jarlskog-88}. One of the six triangles, the
Unitarity Triangle, represents the most experimentally accessible
of the unitarity relations, which involves the two smallest
elements of the CKM matrix, $V_{ub}$ and $V_{td}$. $V_{ub}$ is
involved in $b \to \u$ transitions such as in \B\ meson decays to
charmless final states, and $V_{td}$ appears in $b \to d$
transitions that can proceed via diagrams involving virtual top
quarks, examples of which are the box diagrams describing the
\BzBzb\ mixing. Because the lengths of the sides of the Unitarity
Triangle are of the same order, the angles can be large, leading
to potentially large \CP-violating asymmetries from phases between
CKM matrix elements.

The CKM matrix can be described by four real parameters.  Using
the sine of the Cabibbo angle $\lambda$ as an expansion parameter,
the Wolfenstein parameterization is given
by~\cite{ref:wolfenstein}:
\begin{equation}
  V_{\rm CKM} =
  \begin{pmatrix}
    V_{ud} & V_{us} & V_{ub} \\
    V_{cd} & V_{cs} & V_{cb} \\
    V_{td} & V_{ts} & V_{tb} \
  \end{pmatrix}
  =
  \begin{pmatrix}
    1-\lambda^{2}/2 &  \lambda & A \lambda^{3} (\rho - i\, \eta) \\
    -\lambda & 1-\lambda^{2}/2 & A \lambda^{2} \\
    A \lambda^{3} (1-\rho-i\, \eta) & -A \lambda^{2} & 1 \
  \end{pmatrix}
  + \mathcal{O}(\lambda^{4}) \quad ,
\end{equation}
where $A$, $\bar{\rho}=\rho (1-\lambda^{2}/2)$ and
$\bar{\eta}=\eta (1-\lambda^{2}/2)$ are the remaining three
parameters. It should be noted that, in this parameterization, the
usual (but arbitrary) phase convention under which all the CKM
matrix elements are real except $V_{ub}$ and $V_{td}$ is
implicitly made.

\section{Experimental situation before the \B\ factories}

The sine of the Cabibbo angle $\lambda$ is known at the percent
level ($\lambda \simeq 0.2230$). The parameter $A$, determined
mostly from measurements of semileptonic decays of strange and
beauty particles, is known at the 5\% level ($A \simeq 0.830$).
The coordinates of the apex of the Unitarity Triangle,
$\bar{\rho}$ and $\bar{\eta}$, are constrained by
$|\varepsilon_K|$, $|V_{ub}/V_{cb}|$, \deltambd\ measurements, and
by the limit on $\deltambd / \deltambs$.  These constraints depend
on additional measurements and theoretical inputs.  Many critical
studies of the CKM constraints are available in the recent
literature (see for instance
Ref.~\cite{ref:hoecker}-\cite{ref:bargiotti}); they differ in the
way theoretical uncertainties and experimental systematic errors
are handled, and in the statistical treatment in the combination
of the available information.  The conclusions of the various
analyses are quite consistent however~\cite{ref:beneke_budapest}.
Reference~\cite{ref:hoecker} for instance gives 95\% confidence
intervals $\bar{\rho} \in [ 0.04, 0.38 ]$ and $\bar{\eta} \in [
0.21, 0.49 ]$.   The allowed range for $\bar{\eta}$ is an
experimental indication that the CKM matrix indeed contains a
non-zero phase.

The side of the Unitarity Triangle that is proportional to $V_{td}
V_{tb}^{*}$ forms an angle $\beta$ with the side proportional to
$V_{cd} V_{cb}^{*}$  and an angle $\alpha$ with the side
proportional to $V_{ud} V_{ub}^{*}$.  Experimental sensitivity to
the angles $\beta$ and $\alpha$ can therefore arise from
interferences between the \BzBzb\ mixing amplitude (which involves
$V_{td}$) and decay amplitudes that involve $V_{cb}$ and $V_{ub}$
respectively.  The third angle, $\gamma$, is the argument of
$V_{ub}^*$ in the usual phase convention.   The experimental
programme for testing the CKM model with three generations of
leptons, in particular its description of \CP\ violation in the
charged weak current sector, involves as many measurements of the
sides, angles and other quantities constraining the position of
the apex of the Unitarity Triangle, and checking that the results
are indeed consistent. The \CP-violating observable \stwob\ is
already constrained by the allowed region in
$(\bar{\rho},\bar{\eta})$ from \CP-conserving measurements:
$\stwob \in [0.47, 0.89]$ at the 95\% confidence
level~\cite{ref:hoecker}.

\section{\CP\ violation in the decay}

\CP\ violation in the decay (also referred to as direct \CP
violation) is due to interference among decay amplitudes which
differ in both weak and strong phases.  Direct \CP\ violation is
now firmly established in \KL\ decays: the amount of direct \CP\
violation recently reported by NA48 and KTeV is consistent with
predictions based on the Standard Model within large theoretical
uncertainties~\cite{ref:beneke_budapest}.

For \B\ decays, one builds time-independent \CP\ asymmetry
observables:
\begin{eqnarray}
\label{eq:directCP}
 {\cal A}_{\CP} \equiv \frac{\Gamma( \, \overline{B} \to \overline{f} \, ) - \Gamma(\,
B \to f \, )} {\Gamma(\, \overline{B} \to \overline{f}  \, ) +
\Gamma (\, B \to f \, )} = \frac{ 1 - |
\overline{A}_{\overline{f}} / A_f |^2 }{1 + |
\overline{A}_{\overline{f}} / A_f |^2 }\, .
\end{eqnarray}
Direct \CP\ violation is the only type of \CP\ violation for
charged modes, while for neutral modes it competes with the other
two types of \CP\ violation.  Sizable direct \CP\ violation
effects $(| \overline{A}_{\overline{f}} / A_f | \neq 1)$ require
the contribution to the decay of at least two amplitudes of
comparable size with of course different weak phases, but also a
non-zero relative strong phase (the latter is in general difficult
to estimate theoretically). The rule of thumb is that larger \CP\
violation effects are potentially expected for very rare processes
for which the dominant amplitude ({\it e.g.} a tree amplitude) is
suppressed at the level of higher-order amplitudes ({\it e.g.}
penguin amplitudes).

\begin{table}
\begin{center}
\begin{tabular}{|l|c|c|c||c|}
\cline{2-5}
          \multicolumn{1}{c|}{}             & CLEO~\cite{ref:cleo_pipi} & \babar~\cite{ref:babar_pipi} & \belle~\cite{ref:belle_pipi} & World average \\
\hline
  $\Bz \ \to  \pi^+ \pi^-$ & $4.3^{+1.6}_{-1.4}\pm 0.5$ & $4.1 \pm 1.0 \pm 0.7$ & $5.6^{+2.3}_{-2.0}\pm 0.4$ & $4.44^{+0.89}_{-0.86}$ \\
  $\Bz \ \to  K^+ \pi^-$   & $17.2^{+2.5}_{-1.4}\pm 1.2$ & $16.7 \pm 1.6 \pm 1.3 $ & $19.3^{+3.4}_{-3.2}\, ^{+1.5}_{-0.6}$ & $17.37^{+1.47}_{-1.30}$ \\
  $\Bz \ \to  K^+ K^-$     & $ <1.9 $ & $ <2.5 $ & $ <2.7 $ &  \\
  $B^+  \to  \pi^+ \pi^0$ & $<12.7$ & $5.7^{+2.0}_{-1.8} \pm 0.8$ & $<13.4$ & \\
  $B^+  \to  K^+ \pi^0$   & $11.6^{+3.0}_{-2.7}\, ^{+1.4}_{-1.3}$ & $10.8^{+2.1}_{-1.9}\pm 1.0$ & $16.3^{+3.5}_{-3.3} \, ^{+1.6}_{-1.8}$ & $12.13^{+1.70}_{-1.67}$ \\
  $B^+  \to  K^0 \pi^+$ & $18.2^{+4.6}_{-4.0}\pm 1.6$ & $18.2^{+3.3}_{-3.0}\pm 2.0$ & $13.7^{+5.7}_{-4.8} \, ^{+1.9}_{-1.8}$ & $17.41^{+2.60}_{-2.51}$ \\
  $\Bz \ \to  K^0 \pi^0$ & $14.6^{+5.9}_{-5.1}\, ^{+2.4}_{-3.3}$ & $8.2^{+3.1}_{-2.7}\pm 1.2$ & $16.0^{+7.2}_{-5.6} \, ^{+2.5}_{-2.7}$ & $10.73^{+2.66}_{-2.66}$ \\
\hline
\end{tabular}
\label{tab:charmless} \caption{Recent measurements of branching
ratios for \B\ meson decays to charmless two-body final states
containing pions or kaons, in units of $10^{-6}$.  In certain
cases, 90\% confidence limits are given. }
\end{center}
\end{table}

In charmless hadronic decays for instance, the tree $\b \to \u$
amplitude is highly suppressed and competes with $\b \to \s$ or
$\b \to \d$ penguin amplitudes: asymmetries for these modes could
be as large as ~10\%. The various $\B \to \pi \pi$ and $\B \to K
\pi$  modes have all been recently observed by
CLEO~\cite{ref:cleo_pipi} and confirmed by
\babar~\cite{ref:babar_pipi} and \belle~\cite{ref:belle_pipi} (see
Table~\ref{tab:charmless}), with branching ratios in the
$10^{-6}$-$10^{-5}$ region. \CP\ asymmetries in the self-tagged
modes have been
measured~\cite{ref:babar_pipi}\cite{ref:cleo_cp-asym}\cite{ref:belle_cp-asym}
consistent with zero within errors (see Table~\ref{tab:cp-asym}).
\CP\ asymmetries have also been measured in a variety of other
charmless \B\ decays~\cite{ref:babar_cp-asym} and again no
significant deviation from zero has been observed.  For all these
modes, the sensitivity on ${\cal A}_{\CP}$ is at best at the
10\%-15\% level.

\begin{table}
\begin{center}
\begin{tabular}{|l|c|c|c|}
\cline{2-4}
         \multicolumn{1}{c|}{}             & CLEO~\cite{ref:cleo_cp-asym}
                                           & \babar~\cite{ref:babar_pipi}~\cite{ref:babar_s2a}
                                           & \belle~\cite{ref:belle_cp-asym}  \\
\hline
  $\B \ \to  K^\pm \pi^\mp$   & $ -0.04 \pm 0.16 $  & $ -0.19\pm 0.10$     & $ \ 0.04 ^{+0.19}_{-0.17}$ \\
  $B^\pm  \to  K^\pm \pi^0$   & $ -0.29 \pm 0.23 $  & $ \ 0.00 \pm 0.18 $ & $ -0.06 ^{+0.22}_{-0.20}$ \\
  $B^\pm  \to  K_s^0 \pi^\pm$ & $ \ 0.18 \pm 0.24 $  & $ -0.21 \pm 0.18$    & $ \ 0.10 ^{+0.43}_{-0.34}$ \\
\hline
\end{tabular}
\label{tab:cp-asym} \caption{Recent measurements of charge \CP\
asymmetries in self-tagged $B \to K \pi$ decay modes.}
\end{center}
\end{table}

Is is interesting to look for asymmetries where none is expected.
Examples are the loop-induced $b \to \s \gamma$ modes, for which
${\cal A}_{\CP}$ is strongly suppressed within the Standard Model.
Combining two statistically independent measurements, one based on
the pseudo-reconstruction of the $X_{\s}$ system, the other on
tagging the flavor of the other \B\ with leptons, \cleo\
measures~\cite{ref:cleo_btosg}: ${\cal A}_{\CP}( \b \to \s \gamma
) = -0.079 \pm 0.108 \pm 0.022$. \CP\ asymmetries have also been
measured the exclusive $B^\pm \to K^{*\pm} \gamma$ mode by \cleo\
and \babar~\cite{ref:menke}: ${\cal A}_{\CP}( B^\pm \to K^{*\pm}
\gamma )=+0.08 \pm 0.13 \pm 0.03$ and $-0.035 \pm 0.076 \pm
0.012$, respectively. In both inclusive and exclusive $\b \to \s
\gamma$ analyses, much higher statistics are needed to challenge
the SM on the prediction of very small \CP\ asymmetries.
Similarly, an order of magnitude more data would be needed to test
the prediction of a strong ${\cal A}_{\CP}$ suppression in pure
penguin $b \to \s \sbar \s$ decays, such as the recently observed
$B \to \phi K$
modes~\cite{ref:cleo_phik}\cite{ref:babar_phik}\cite{ref:belle_phik}.

The copious $\b \to \c \cbar s$ decays are in general dominated by
the tree amplitude. In modes such as $B^{\pm} \to \jpsi K^{\pm}$,
the tree amplitude is colored-suppressed but the dominant penguin
contribution has nearly the same weak phase and direct \CP
violation further suppressed. This is experimentally confirmed at
the 4\% level by \cleo~\cite{ref:cleo_jpsik}: ${\cal A}_{\CP}(
B^{\pm} \to \jpsi K^{\pm} ) = (+1.8 \pm 4.3 \pm 0.4 )\%$ and at
the 3\% level by \babar~\cite{ref:menke}: ${\cal A}_{\CP}( B^{\pm}
\to \jpsi K^{\pm} ) = (-0.9 \pm 2.7 \pm 0.5 )\%$.

\section{\CP\ violation in mixing}

\CP\ (or \T) violation in  \BzBzb\ mixing (also referred to as
indirect \CP\ violation) manifests itself as an asymmetry in the
transitions $\Bz \to \Bzb$ and $\Bzb \to \Bz$, as a consequence of
the mass eigenstates being different from the \CP\ eigenstates:
\begin{equation}
\label{eq:indirect} \left| \, q / p \,  \right| \neq 1  \, \,
\Longrightarrow \, \, \textrm{Prob}( \Bz_{phys}(t) \to \Bzb )\neq
\textrm{Prob}( \Bzb_{phys}(t) \to \Bz )  \, . \nonumber
\end{equation}
This effect can be studied by investigating time-dependent
differences in mixing rates in decays to flavor-specific final
states such as semileptonic neutral \B\ decays:
\begin{equation}
\label{eq:AT}
 {\cal A}_T(t) = \frac{\Gamma( \, | \Bzb_{\rm {phys}}(t)
\rangle \to \ell^+ \nu X \, ) - \Gamma(\,  | \Bz_{\rm {phys}}(t)
\rangle \to \ell^- \nu X \, )} {\Gamma(\,  | \Bzb_{\rm {phys}}(t)
\rangle \to \ell^+ \nu X \, ) + \Gamma (\,  | \Bz_{\rm {phys}}(t)
\rangle \to \ell^- \nu X \, )} \, .
\end{equation}
Starting from equations~\ref{eq:timeEvolution}, one finds that the
proper time dependence cancels out in the ratio~\ref{eq:AT} and
the asymmetry is independent of $t$:
\begin{equation}
\label{eq:AT_2} {\cal A}_T(t) = a_T \ \ \ \ {\rm with} \ \ \ \
 a_T \equiv \frac{1-|q/p|^{4}}{1+|q/p|^{4}}
\,.
\end{equation}
The \CP\ asymmetry parameter $a_T$ can be extracted from both
time-integrated and time-dependent measurements. At LEP and the
Tevatron, time-dependent asymmetries are studied using either
flavor-tagged samples of semileptonic decays or fully inclusive
samples of \Bz\ decays.  In the latter case, the time-integrated
rate vanishes due to \CPT\ symmetry but some sensitivity to $a_T$
exists in the time-dependence of the asymmetry,  as shown in
Ref.~\cite{ref:beneke-buchalla-dunietz}. Averaging the results of
their various analyses,  CDF~\cite{ref:cdf_at},
OPAL~\cite{ref:opal_at99} and ALEPH~\cite{ref:aleph_at} obtain
$a_T = 0.024 \pm 0.063 \pm 0.033$, $a_T = 0.004 \pm 0.056 \pm
0.012$ and $a_T = -0.013 \pm 0.026 ({\rm stat+syst})$,
respectively. At the \FourS, CLEO has recently measured the
integrated like-sign dilepton charge asymmetry~\cite{ref:cleo_at};
the result is in agreement with its previous measurement of $a_T$
via partial hadronic reconstruction. The weighted average of the
two CLEO measurements gives: $a_T = 0.014 \pm 0.041 \pm 0.006$. At
this Conference, \babar\ has presented a time-dependent analysis
of its like-sign dilepton sample based on  20.7\invfb\ of
data~\cite{ref:babar_at}, obtaining: $ a_T = 0.0048 \pm 0.0116 \pm
0.0144$. This preliminary result  demonstrates that asymmetric \B\
factory experiments are already close to sensitivities needed to
test theoretical predictions on indirect \CP\ violation ($a_T \leq
 10^{-2}$), and that systematic uncertainties can be controlled at
the required level.

To conclude, no significant indirect \CP\ violation effect in the
neutral \B\ meson system has been seen to date. This experimental
constraint allows us to express, to a very good approximation, the
ratio $q/p$ in term of a pure phase $\phi_M$: $q/p =
e^{2i\phi_M}e^{2i\xi_B}$ ($\xi_B$ is maintained to express the
arbitrariness of the phase convention). The phase $\phi_M$ results
from CKM factors involved in the box diagrams that describe the
dispersive part of the $\Bz \to \Bzb$ amplitude. In the Standard
Model, to a very good approximation:
\begin{equation}
  \frac{q}{p} = \frac{  V_{td} V_{tb}^* }{V_{td}^* V_{tb} } e^{2i\xi_B}
\end{equation}
and therefore $ \phi_M =  \arg ( V_{td} V_{tb}^* ) = \pi - \beta +
\arg ( V_{cd} V_{cb}^* ) $, which, in the usual phase convention,
reduces to $\phi_M = -\beta$. New Physics can possibly manifest
itself as a phase shift with respect to the Standard Model
expectation for $\phi_M$.

\section{\CP\ violation in interference between decay with/without mixing}

In order to calculate the time-dependent decay rates to a specific
final state $f$ accessible to both \Bz\ and \Bzb\ decays, one
introduces the phase-independent complex parameter $\lambda_f$:
\begin{equation}
 \lambda_f \equiv \frac{q}{p} \frac{\overline{A}_f}{A_f} \, ,
\end{equation}
where $ A_f \equiv \langle f | H | \Bz \rangle $ and $
\overline{A}_f \equiv \langle f | H | \Bzb \rangle $. Particularly
interesting is the situation where $| A_f |^2 \approx |
\overline{A}_f |^2$. This condition is fulfilled  in the case that
we consider now, where the final state $f$ is a \CP\ eigenstate,
$f_{\CP}$. Using equations~\ref{eq:timeEvolution}, one obtains
\begin{equation}\label{eq:decayRateB0}
  {\rho}_\pm(t) =
  \left\{
  \frac{1}{2} \left( \, 1+ |\lambda_{f_{\CP}}|^2 \right)
  \pm \frac{1}{2} \left( \, 1 -  |\lambda_{f_{\CP}}|^2 \right) \cos{(\deltambd  t)}
  \mp \textrm{Im}(\lambda_{f_{\CP}}) \sin{(\deltambd  t)}
   \right\}
   e^{-{\rm \Gamma\, t}} \, ,
\end{equation}
with $\rho_+(t) \equiv | \langle f_{\CP} | H | \Bz_{\rm {phys}}(t)
\rangle | ^2 / \, | A_{f_{\CP}} |^2 $ and $\rho_-(t) \equiv |
\langle f_{\CP} | H | \Bzb_{\rm {phys}}(t) \rangle | ^2 / \, |
\overline{A}_{f_{\CP}} |^2 $.  One finds that the conditions for
no \CP\ violation at any time $t$ are $| \lambda_{f_{\CP}} | = 1 $
and $\textrm{Im}\, \lambda_{f_{\CP}} =  0$:
\begin{equation}
\lambda_{f_{\CP}} \neq \pm 1 \  \Longrightarrow  \ \textrm{Prob}(
\Bz_{phys}(t) \to f_{\CP}\, )\neq \textrm{Prob}( \Bzb_{phys}(t)
\to f_{\CP}\, ) \, .
\end{equation}
The time-dependent \CP\ asymmetry:
\begin{equation}
\label{eq:ACP}
 {\cal A}_{f_{\CP}}(t) \equiv \frac{\Gamma( \, | \Bzb_{\rm {phys}}(t)
\rangle \to f_{\CP}\, ) - \Gamma(\,  | \Bz_{\rm {phys}}(t) \rangle
\to f_{\CP}\, )} {\Gamma(\,  | \Bzb_{\rm {phys}}(t) \rangle \to
f_{\CP}\, ) + \Gamma (\, | \Bz_{\rm {phys}}(t) \rangle \to
f_{\CP}\, )}
\end{equation}
can be written as:
\begin{equation}
{\cal A}_{f_{\CP}}(t) = S_{f_{\CP}} \cdot \sin{(\deltambd  t)} -
C_{f_{\CP}} \cdot \cos{(\deltambd  t)} \label{eq:asymmetry} \, ,
\end{equation}
where the coefficients of the sine and cosine terms are:
\begin{equation}
S_{f_{\CP}} = \frac{ 2\, \textrm{Im} \lambda_{f_{\CP}} }{\, 1+
|\lambda_{f_{\CP}}|^2 } \ \ \
 \textrm{and}\ \ \
C_{f_{\CP}} =  \frac{\,  1 - |\lambda_{f_{\CP}}|^2 }{\, 1 +
|\lambda_{f_{\CP}}|^2 } \, .
\end{equation}
The cosine term vanishes in absence of both \CP\ violation in
mixing ($|q/p|=1$) and direct \CP\ violation in the decay ($ |
\AbfCP / \AfCP | = 1$).  Even in that case,  \CP\ violation can
arise from the weak phase difference between $q/p$ and
$\AbfCP/\AfCP$, resulting in a non-vanishing sine term
($\textrm{Im}\, \lambda_{f_{\CP}}\neq 0$). In the language of
flavor eigenstates, this is interpreted as an interference between
the decay of a \Bz\ meson with and without mixing \footnote{note
that, if the interpretation of this \CP-violating effect is
somewhat convention-dependent, the phase of $\lambda_{f_{\CP}}$ is
not.}.

In the case of  a single \CP-violating phase in the disintegration
process, $\phi_D$,  one finds:
\begin{equation}
  \frac{ \overline{A}_{f_{\CP}} } { A_{f_{\CP}} } = \eta_{f_{\CP}} \, e^{2i\phi_D}\,
  e^{-2i\xi_B} \, ,
\end{equation}
where $\eta_{f_{\CP}}$ is the \CP\ parity of the $f_{\CP}$ final
state, $\eta_{f_{\CP}}=\pm1$.  It follows that:
\begin{equation}
\label{eq:Sfcp_noDirectCP} \lambda_{f_{\CP}}= \eta_{f_{\CP}}\,
e^{2i(\phi_D+\phi_M)}, \ \ \ S_{f_{\CP}} = \textrm{Im} \,
\lambda_{f_{\CP}} = \eta_{f_{\CP}}\, \sin{ 2(\phi_D+\phi_M) } , \
\ \ C_{f_{\CP}}=0,
\end{equation}
and
\begin{equation}
 {\cal A}_{f_{\CP}}(t) = \eta_{f_{\CP}}\, \sin{2(\phi_D+\phi_M)}\, \sin{(\deltambd
 t)}\ .
\end{equation}

The "gold-plated" $\jpsi K^0_{S,L}$ modes provide a theoretically
clean way of extracting \stwob:  the branching ratio are
relatively high ($\sim 10^{-4}$) and the experimental signatures
are clear.  The quark subprocess at the tree level for the $\Bzb
\to \jpsi \Kzb$ is $\b \to \c \cbar s$, which involves the
$V_{cs}^*V_{cb}$ product of CKM factors. The interference in $\Bz
(\Bzb) \to \jpsi K^0_{S,L}$ is possible due to $\KzKzb$ mixing
which introduces another CKM product, $V_{cd}^*V_{cs}$. With
negligible theoretical uncertainty, direct \CP\ violation is ruled
out in these modes (this is supported experimentally by the
non-observation of direct \CP\ violation in the SU(3)-related
$B^{\pm} \to \jpsi K^{\pm}$ decay). In the usual phase convention,
the ratio of amplitudes is made real, $\phi_D=0$, and all the
effect comes from the phase $\phi_M=-\beta$ in $q/p$, {\it i.e.}
from the fact that the mass eigenstates are not \CP-even and
\CP-odd. One finds:
\begin{equation} \label{eq:cp_asym}
{\cal A}_{\jpsi K^0_{S,L}} = - \eta_{\jpsi K^0_{S,L}} \, \sin{ 2
\beta } \, \sin{ ( \deltambd t ) } \, ,
\end{equation}
where $\eta_{\jpsi K^0_{S}}$ and  $\eta_{\jpsi K^0_{L}}$ are equal
to $-1$ and $+1$, respectively.  At LEP, OPAL~\cite{ref:opal_s2b}
with 24 $\Bz \to \jpsi \KS$ candidates (60\% purity) selected out
of 4.4 million hadronic $Z^0$ decays measures: $\stwob = 3.2
{^{+1.8}_{-2.0}}{\rm (stat)} \pm 0.5{\rm (syst)}$.
ALEPH~\cite{ref:aleph_s2b} with 23 candidates (71\% purity)
selected out of 4 million hadronic $Z^0$ decays measures: $\stwob
= 0.84 {^{+0.82}_{-1.04}}{\rm (stat)} \pm 0.16{\rm (syst)}$. At
the Tevatron (FNAL), CDF~\cite{ref:cdf_s2b} using $~400$ $\Bz \to
\jpsi \KS$ events out of the entire Run~I data sample (110\invpb\
at $\sqrt{s}=1.8\tev$) measures: $ \stwob = 0.79
{^{+0.41}_{-0.44}}{\rm (stat+syst)}$.

\subsection{The concept of an asymmetric \B\ factory}

The cleanest source of \B\ mesons is in \epem\ collisions at the
energy of the \FourS\ resonance ($\sqrt{s}\approx 10.58\gev$). The
\FourS\ resonance is the lightest $\Upsilon$ resonance ($\b\bbar$
bound state with $J^{PC}=1^{--}$) above \B\ meson pair production
threshold. To present knowledge, the \FourS\ decays exclusively
into \BpBm\ or \BzBzb\ final states in equal amounts.  The
$\epem\to\FourS$ process, with a cross-section close to 1\nb,
competes with the QED pair production of light quarks
$\epem\to\qqbar$ (where $q=$\u, \d, \s\ or \c). The background
from QED continuum constitutes 75\% of the hadronic cross-section,
but can be reduced thanks to distinct topological characteristics.
Furthermore, the QED continuum can be studied with data taken at
energies under the \BB\ threshold, typically 40\mev\ below the
\FourS\ resonance ({\it off-resonance} data).

Due to the intrinsic spin of the \FourS, \BB\ states produced in
the $\FourS\to\BB$ reaction are in a coherent $L=1$ state. In the
case of the $\FourS\to\BzBzb$ reaction, each meson evolves
according to the time evolution of single \B\ meson
equations~\ref{eq:timeEvolution}. However, the two mesons evolve
in phase, and the correlation between both sides of the \BzBzb\
system holds at any time after production until one of the two
mesons decays. If the first meson decays into a flavor specific
decay mode, the other meson in the pair, at that same instant,
must have the opposite flavor.  If the first meson decays into a
\CP\ eigenstate, the other meson in the pair, at that same
instant, must have opposite \CP.

Let us consider the case where one of the two mesons (called
$B_{\rm tag}$) decays into a flavor-specific final state at proper
time $t_{\rm tag}$, while the other meson (called $B_{\rm rec}$,
because it is usually fully-reconstructed) decays into the final
state $f$ at proper time $t_{\rm rec}$. After integration on
$t_{\rm rec}+t_{\rm tag}$, the decay rate writes:
\begin{eqnarray} \label{eq:decayRate}
  {\rho}(\Delta t, \varepsilon_{\rm tag}) \ \propto \ e^{-| \Delta t | / \tau_{B_d} } \,   \cdot
  \{ \ \, ( \, |A_f|^2 + |\overline{A}_f|^2 \, ) \ \ \ \ \ \ \ \ \ \ \ \ \ \ \ \ \ \ \ \ \ \ \ \ \ \ \ \ \ \ \ \ \ \ \ \ \ \ \ \ \ \ \ \   \nonumber \\
   \    + \ \varepsilon_{\rm tag} \, [ \ 2 \ {\rm Im}(\, \frac{q}{p}  A^{*}_f
\overline{A}_f \, ) \, \sin{ ( \deltambd \Delta
  t) }  \, -  \, ( \, |A_f|^2 - |\overline{A}_f|^2 \, ) \, \cos{ ( \deltambd \Delta t )
  } \ ] \ \} \ ,
\end{eqnarray}
where  $\Delta t = t_{\rm rec} - t_{\rm tag}$ is the proper decay
time difference and $\varepsilon_{\rm tag}$ equals $1$ or $-1$
depending or whether the $B_{\rm tag}$ is identified as a \Bz\ or
a \Bzb.

If the final state $f$ is also a flavor-specific final state
($B_{\rm rec}=B_{\rm flav}$), one obtains from the above equation
the normalized \BzBzb\ mixing time distribution:
\begin{eqnarray}
h(\, \Delta t, \, \varepsilon_{\rm tag} \times \varepsilon_f ) =
 \frac{1}{4 \tau_{B_d}} \, e^{- | \Delta t |/ \tau_{B_d} } \, \{
\, 1 \, - \, \varepsilon_{\rm tag} \times \varepsilon_f \, \cdot
\cos{ ( \deltambd \Delta t ) } \, \} \ , \label{eq:pdf_mixing}
\end {eqnarray}
where $\varepsilon_f = +1$ or $-1$ depending on whether $f$ is a
\Bz\ decay final state ($| \overline{A}_f | = 0 $) or  a \Bzb\
decay final state ($| A_f | = 0 $). The cases $\varepsilon_{\rm
tag} \times \varepsilon_f = +1 $ and $\varepsilon_{\rm tag} \times
\varepsilon_f = -1$ define the mixed and the unmixed samples,
respectively. (If one considers the mixed and unmixed samples
together, one obtains a pure lifetime distribution.) Similarly, if
the final state $f$ is a \CP\ eigenstate ($B_{\rm rec}=B_{\CP}$),
the normalized decay distributions writes:
\begin{equation}
f( \, \Delta t, \, \varepsilon_{\rm tag} ) = \frac{1}{4
\tau_{B_d}} e^{- | \Delta t |/ \tau_{B_d} }  \left\{ 1 +
\varepsilon_{\rm tag}  \left( S_{f_{\CP}} \cdot \sin{\deltambd
\Delta t}  -  C_{f_{\CP}} \cdot \cos{\deltambd \Delta t}  \right)
 \right\} \  .
 \label{eq:pdf_cp}
\end{equation}
The time distributions at the \FourS\ are therefore obtained by
substituting  the proper time difference $\Delta t$ to the \Bz\
decay time $t$, with the difference that $\Delta t$ is an
algebraic quantity which takes values from $-\infty$ to $+\infty$.
When integrating over $\Delta t$, one loses the information on the
coefficient of the sine term. As a result of the coherent
production of \Bz\ mesons at the \FourS, a time-integrated \CP\
asymmetry measurement is insensitive to ${\rm
Im}(\lambda_{f_{\CP}})$; the study the \CP\ asymmetry as a
function of $\Delta t$ is required. Experimentally, the variable
$\Delta t$ can be related to the distance between the locations of
the two \B\ meson decays.  In practice, at an energy-symmetric
machine like CESR, the flight distance of the \B\ mesons
($\sim30\mum$) is too small compared to the interaction region
size, and such a time-dependent study is impossible even with
perfect vertex resolution.

\begin{figure}[htb]
\begin{center}
\epsfig{file=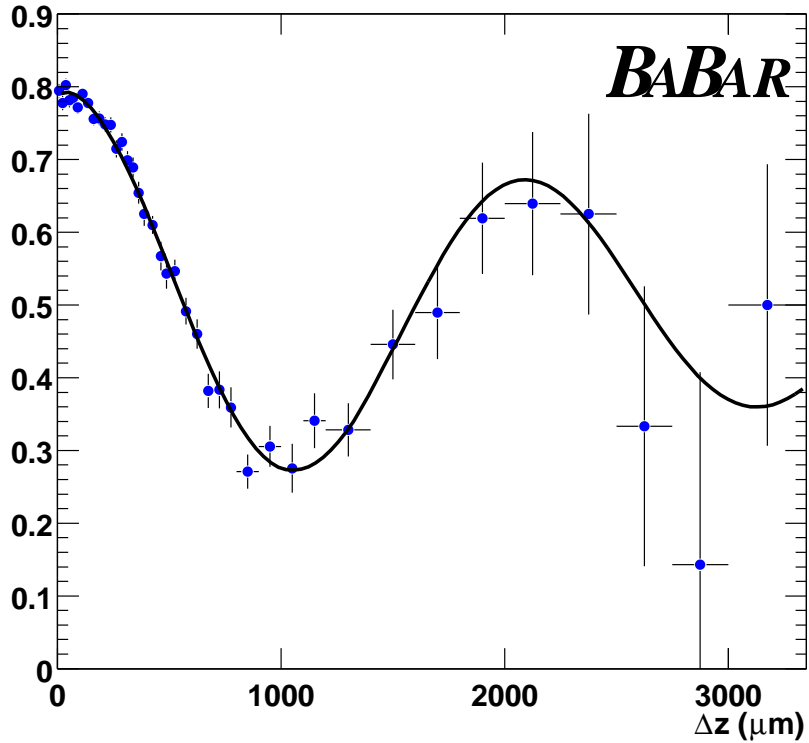, width=12cm} \caption{ Time-dependent
relative rate of unlike-sign dilepton events (\babar\ data,
$20.7\invfb$) and the overlaid binned maximum likelihood fit from
which the \BzBzb\ oscillation frequency $\deltamd$ is
measured~\cite{ref:touramanis}: $\deltambd = 0.499 \pm 0.010 \pm
0.012 \, \hbar \ps^{-1}$. The full scale, expressed here in \mum,
covers more than 15 lifetimes of the \Bz\ meson (about $1.5$
oscillation period). \label{fig:babar_dilepton}}
\end{center}
\end{figure}

In the late 1980s, to resolve this problem, the concept of a
machine operating at the \FourS\ in an asymmetric mode, {\it i.e.}
with beams of unequal energies, was proposed~\cite{ref:oddone}.
\pepii\ and \kekb\ were built on this model. At asymmetric \B\
factories, the \FourS\ is produced in motion in the laboratory
frame, and the two \B\ mesons travel measurable distances along
the boost axis before they decay. The relation between the proper
time difference $\Delta t$ and the distance between the two \B\
decay vertices along the boost axis $\Delta z$ is (to a very good
approximation) $\Delta z = \beta \gamma c \Delta t$. The
center-of-mass frame boosts for  \pepii\ and \kekb\ are similar:
$\beta \gamma = 0.56$ and $\beta \gamma = 0.45$, respectively. One
lifetime of the \Bz\ meson corresponds to a distance of the order
of $250\mum$ along the boost axis and a complete \Bz\Bzb\ mixing
period $2 \pi / \deltambd$  to about $2\mm$. (This is illustrated
on  Fig.~\ref{fig:babar_dilepton}, which shows the time-dependent
$\ell^\pm\ell^\mp$ relative rate for dilepton events in \babar\
data  as a function of $\Delta z$.) Time-dependent \CP\
asymmetries are expected to be maximal when mixed and unmixed
samples are of equal size, around a quarter period ($\sim
500\mum$).

\subsection{Mis-tagging probability and time-resolution function}

Flavor tagging of selected events requires the determination of
$\varepsilon_{\rm tag}$, the flavor of the $B_{\rm
tag}$~\footnote{thanks to the coherent evolution of \Bz\ mesons at
the \FourS, tagging the flavor of the $B_{\rm tag}$ when it decays
is equivalent to tagging the flavor of the $B_{\rm rec}$ at
$\Delta t = 0$.}. In practice, of course, flavor tagging is
imperfect. Let \mistag\ be the probability for a signal event to
be incorrectly assigned a flavor tag $\varepsilon_{\rm tag}$. The
perfect time-dependent probability density for this event $\rho
(\Delta t, \varepsilon_{\rm tag})$ has to be replaced by $( 1 -
\mistag ) \times \rho ( \Delta t, \varepsilon_{\rm tag}  ) +
\mistag \times \rho ( \Delta t, -\varepsilon_{\rm tag} ) = \rho (
\Delta t, ( 1 - 2\mistag  ) \times \varepsilon_{\rm tag} ) $,
where we use the fact that $\rho$ is linear in $\varepsilon_{\rm
tag}$. The effect of the imperfect flavor tagging is therefore to
replace $\varepsilon_{\rm tag}$ by ${\cal D} \times
\varepsilon_{\rm tag}$ in front of oscillatory terms in
Eq.~\ref{eq:decayRate}-~\ref{eq:pdf_cp}, where ${\cal D}\equiv
1-2\mistag$ is a dilution factor due to mis-tagging.

Flavor tagging  is principally based on charge correlations of
daughter particles with the flavor of the decaying $\B_{\rm tag}$.
For example, the presence of the following particles would
identify a $\Bzb$ decay: high-momentum $\ell^-$ leptons ($e^-$ and
$\mu^-$) from $\b \to \c \ell^- \nu$ semileptonic decays;
intermediate-momentum $\ell^+$ lepton from cascade decays; $K^+$
from charm decays; high-momentum $\pi^-$ from $\Bzb \to
D^{*+}\pi^-$ decays; soft $\pi^+$ from $D^{*+} \to \Dz \pi^+$
decays.  \babar\ and \belle\ use different strategies to combine
this information.  \babar\ defines four tagging categories based
on the physical content of the event (such as the presence of a
lepton or a kaon) while \belle\ uses a multi-dimensional
likelihood method and defines six tagging categories based on the
value of a continuous flavor tagging dilution factor.  The tagging
performances of the two experiments are very similar, with an
effective tagging efficiency $Q\approx 27\%$ (this means that a
sample of 100 signal events is statistically equivalent to a
sample of 27 perfectly tagged events).

The vertex resolution in $z$ for the fully-reconstructed $B_{\rm
rec}$ is of the order of 60\mum, depending on the mode.  The
position of the vertex of the $B_{\rm tag}$ is determined using
the charged tracks not belonging to the $B_{\rm rec}$, and by
exploiting energy-momentum conservation and the knowledge of the
beam spot position. A precision on $\Delta z$ of the order of
180\mum is obtained (this is to be compared to the average flight
distance $\sim250\mum$). The $\Delta z$ determination algorithms
have high efficiencies, typically greater than 97\%.

The final probability density functions $\varrho(\Delta t)$ are
obtained by convoluting the ideal time distributions $\rho(\Delta
t)$ with a time resolution function ${\cal R}(\delta_{\Delta t})$:
\begin{eqnarray}
\varrho(\Delta t) \equiv (\rho \otimes {\cal R} )( \Delta t ) =
\int_{-\infty}^{+\infty} \rho( \Delta t ) \times {\cal R}(
\delta_{\Delta t} ) \cdot {\rm d}\delta_{\Delta t} \ ,
\end{eqnarray}
where $\delta_{\Delta t} = \Delta t - \Delta t_{\rm true}$.  The
time resolution functions are parameterized as a normalized sum of
two (\belle) or three (\babar) Gaussian distributions with
different means and widths.  (In \babar, the third Gaussian has a
fixed very large width and no offset, and accounts for fewer than
1\% of events with badly reconstructed vertices.) In both
analyses, the core and tail distributions are allowed to have
non-zero means to account for the bias due to the flight of charm
particles whose decay products are used to determine the position
of the $B_{\rm tag}$ decay vertex, and their widths are scaled by
event-by-event errors derived from the vertex fits.  In \babar,
the parameters of the resolution function are evaluated
independently for each tagging category.

\begin{figure}[htb]
\begin{center}
\epsfig{file=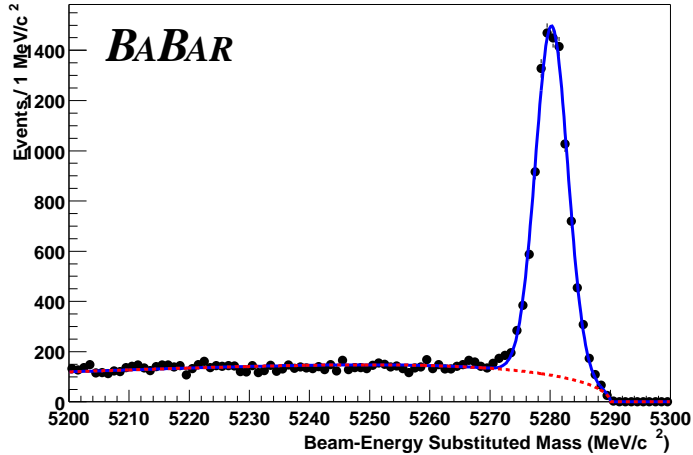, width=10cm}
\caption{ Mass distribution for
\babar's sample of flavor-specific fully-reconstructed hadronic
\Bz\ decays, called the $B_{\rm flav}$ sample in the text. (The
mass is computed with the beam energy substituted to the measured
energy of the candidates.) This sample, composed of $~9360$ signal
events selected out of 29.1\invfb\ of data, is used for lifetime
and mixing measurements (see text).
\label{fig:babar_flavor-sample}}
\end{center}
\end{figure}

\subsection{Flavor and \CP\ samples}

The flavor $B_{\rm flav}$ sample
(Fig.~\ref{fig:babar_flavor-sample}) is composed of events where
the final state $f$ is a flavor-specific hadronic state, such as
$\Bz \to D^{(*)-} \pi^+$, $D^{(*)-} \rho^+$, $D^{(*)-} a_1^+$,
$\jpsi K^{*0}(\to K^+\pi^-)$ and charge conjugates. (Similar decay
modes are considered for charged \B\ mesons.)  The selection rate
for these Cabbibo-favored modes is of the order of 300 events per
\invfb, with purities around $85\%$.

\begin{figure}[htb]
\begin{center}
\epsfig{file=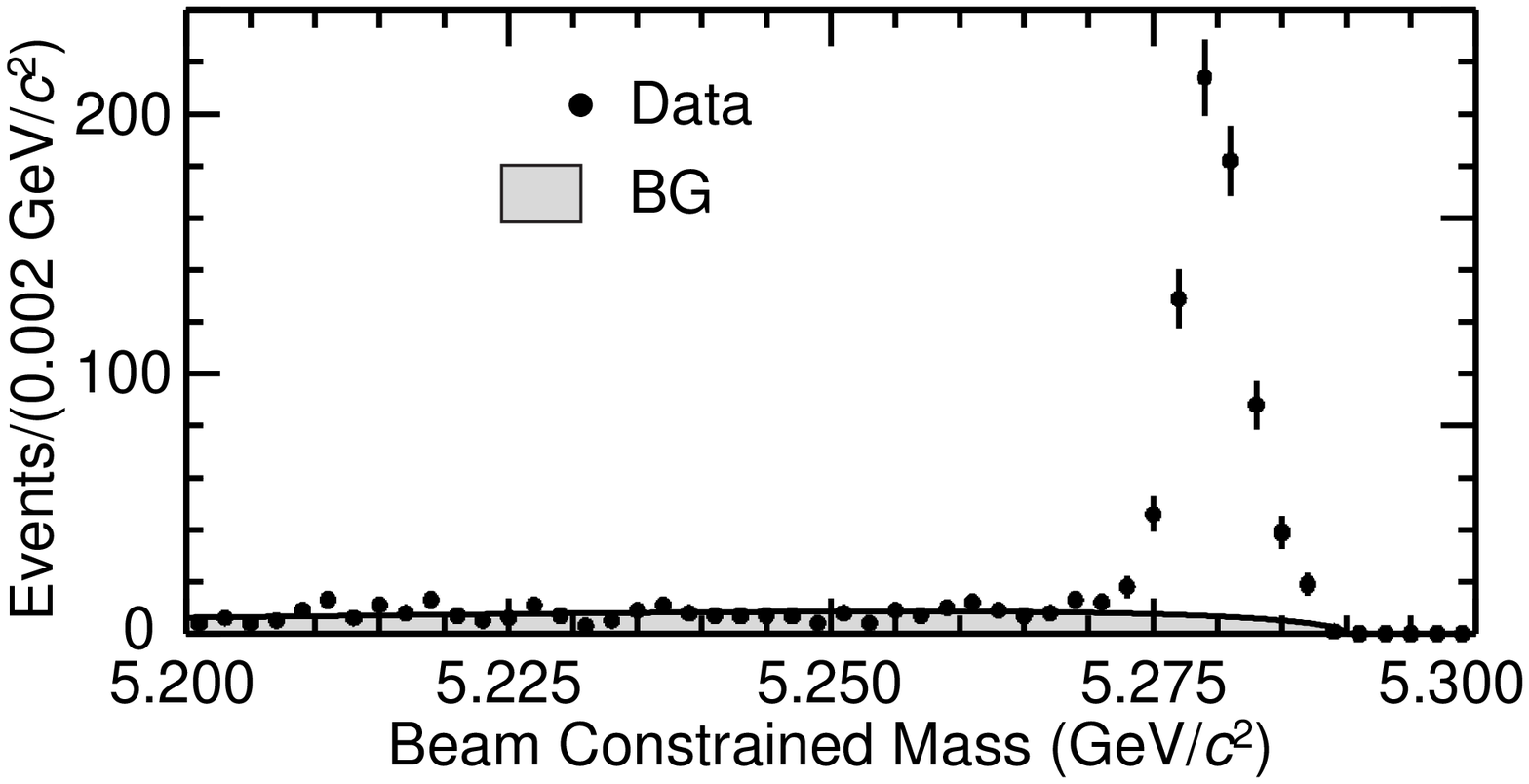, width=10cm} \caption{ Beam
energy-constrained mass distribution for \belle's high-purity
sample of $\eta_{\CP}=-1$ events for the \stwob\ fit (706 events
selected out of 30\invfb\ of data). \label{fig:belle_cp-sample}}
\end{center}
\end{figure}

The \CP\ sample  is divided into three categories:
$\eta_{\CP}=-1$, $\eta_{\CP}=+1$ and mixed-\CP. \belle\ uses all
selected events  for the \CP\ sample while \babar\ applies further
cuts, rejecting 30\% of events with poor or very poor flavor
tagging information. The $\eta_{\CP}=-1$ sample
(Fig.~\ref{fig:belle_cp-sample}) is composed of events selected in
the $\Bz \to \jpsi \KS$, $\psitwos \KS$ and $\chi_{c1} \KS$ modes.
Are considered the decays: $\jpsi \to \ell^+\ell^-$; $\KS \to
\pi^+ \pi^-$ (and in certain cases $\pi^0 \pi^0$); $\psitwos \to
\ell^+\ell^-$ and $\jpsi \pi^+ \pi^-$; $\chi_{c1} \to \jpsi
\gamma$. \belle\ also considers $\Bz \to \eta_c \KS$ decays, with
$\eta_c \to \KS K \pi$ and $K^+ K^- \pi^0$.  \belle\ selects 706
events with 93\% purity, while \babar\  keeps 480 events with 96\%
purity after tagging cuts. The $\eta_{\CP}=+1$ sample is composed
of events selected in the $\Bz \to \jpsi \KL$ mode: 569 events
with 60\% purity for \belle\ and 273 events for 51\% purity after
tagging cuts for \babar. The mixed-\CP\ sample is composed of
events selected in the $B^0 \to \jpsi K^{*0}$ mode with $K^{*0}
\to \KS \pi^0$: 41 events with 84\% purity for \belle\ and 50
events with 74\% purity after tagging cuts for \babar.  For this
$VV$ mode, the \CP\ content can be extracted from an analysis of
angular distributions of the related non-\CP\ $\jpsi K^*$ modes.
\babar~\cite{ref:babar_psikstar} and
\belle~\cite{ref:belle_psikstar} each use their own values of
$R_T$, the fraction of \CP-odd in the decay.  The results,  $R_T=
0.16 \pm 0.03 \pm 0.03$ and $R_T = 0.19 \pm 0.04 \pm 0.04$,
respectively, are consistent with less-precise previous
measurements.

The information on the interesting physical quantities (for
instance $\tau_{B_d}$, $\deltambd$ or $\stwob$) is extracted from
un-binned maximum likelihood fits to the $\Delta t$ spectra of
either flavor or \CP\ samples (or a combination of the two
samples) using the probability density functions defined above for
the signal. The fits also include empirical descriptions of
background $\Delta t$ distributions, the parameters of which are
determined mostly from events in the sidebands of the signal
region or from events selected in off-resonance data. $\tau_{B_d}$
and \deltambd\ are fixed to their PDG value~\cite{ref:pdg} in the
fits for \stwob.  The mis-tagging probabilities and the parameters
of the resolution function for the signal and the background are
fit parameters. \babar\ also fits for a possible difference
between \Bz\ and \Bzb\ tags. For the fit to the $\jpsi K^*$
sample, \belle\ includes the event-by-event information of the
$K^*$ helicity angle in the \CP\ fit.  \babar\ performs a global
fit to the flavor and \CP\ sample in order to get a correct
estimate for the errors and their correlations: the largest
correlation between \stwob\ and any combination of the other
fitted parameters is found to be as small as 13\%.

\subsection{Observation of \CP\ violation in the \B\ meson system}

\begin{figure}[htb]
\begin{center}
\epsfig{file=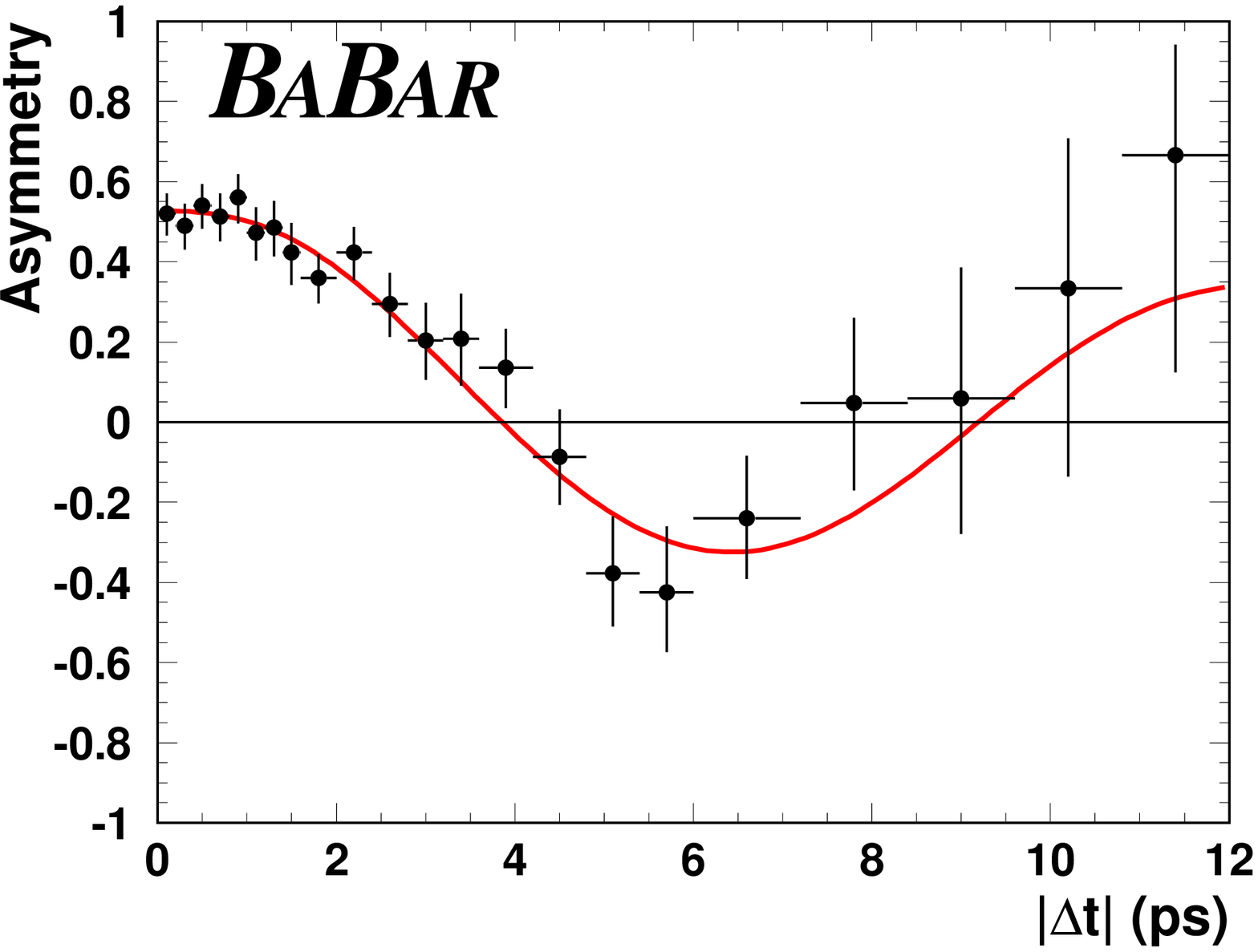, width=15cm} \caption{ Time-dependent
asymmetry for the flavor sample (\babar, $20.7\invfb$), with,
superimposed, the projection of the un-binned maximum likelihood
fit, from which a competitive measurement of the \BzBzb\ mixing
frequency is extracted (see text).  This sample is included in the
global fit for \stwob\ and dominates the determination of  the
mis-tagging probabilities and parameters of the time resolution
function. \label{fig:babar_mixing-asymmetry}}
\end{center}
\end{figure}

Lifetimes measurements with fully-reconstructed decays are
performed using the first two building blocks of the \stwob\
analysis: the reconstruction of \B\ mesons in flavor eigenstates,
and the $\Delta z$ determination. Using its $B_{\rm flav}$ and
charged \B\ samples, \babar\ has produced precision measurements
of \Bz\ and $B^\pm$ lifetimes and their ratio at the 2\%
level~\cite{ref:babar_lifetimes}: $\tau_{\Bz} = 1.546 \pm 0.032
\pm 0.022 \ps$, $\tau_{B^\pm} = 1.673 \pm 0.032 \pm 0.023 \ps$ and
$\tau_{B^\pm}/\tau_{\Bz} = 1.082 \pm 0.026 \pm 0.012$. \belle\ has
lifetime results using $B \to D^* \ell \nu$ semileptonic decays
with comparable precision~\cite{ref:schrenk}: $\tau_{\Bz} = 1.55
\pm 0.02 \ps$, $\tau_{B^\pm} = 1.64 \pm 0.03 \ps$. Mixing
measurements use in addition the third building block which is
flavor tagging.  Using its $B_{\rm flav}$ sample,
\babar~\cite{ref:babar_mixing} measures $\deltambd = 0.519 \pm
0.020 \pm 0.016 \, \hbar \ps^{-1}$ (see
Fig.~\ref{fig:babar_mixing-asymmetry}).

\begin{figure}[htb]
\begin{center}
\epsfig{file=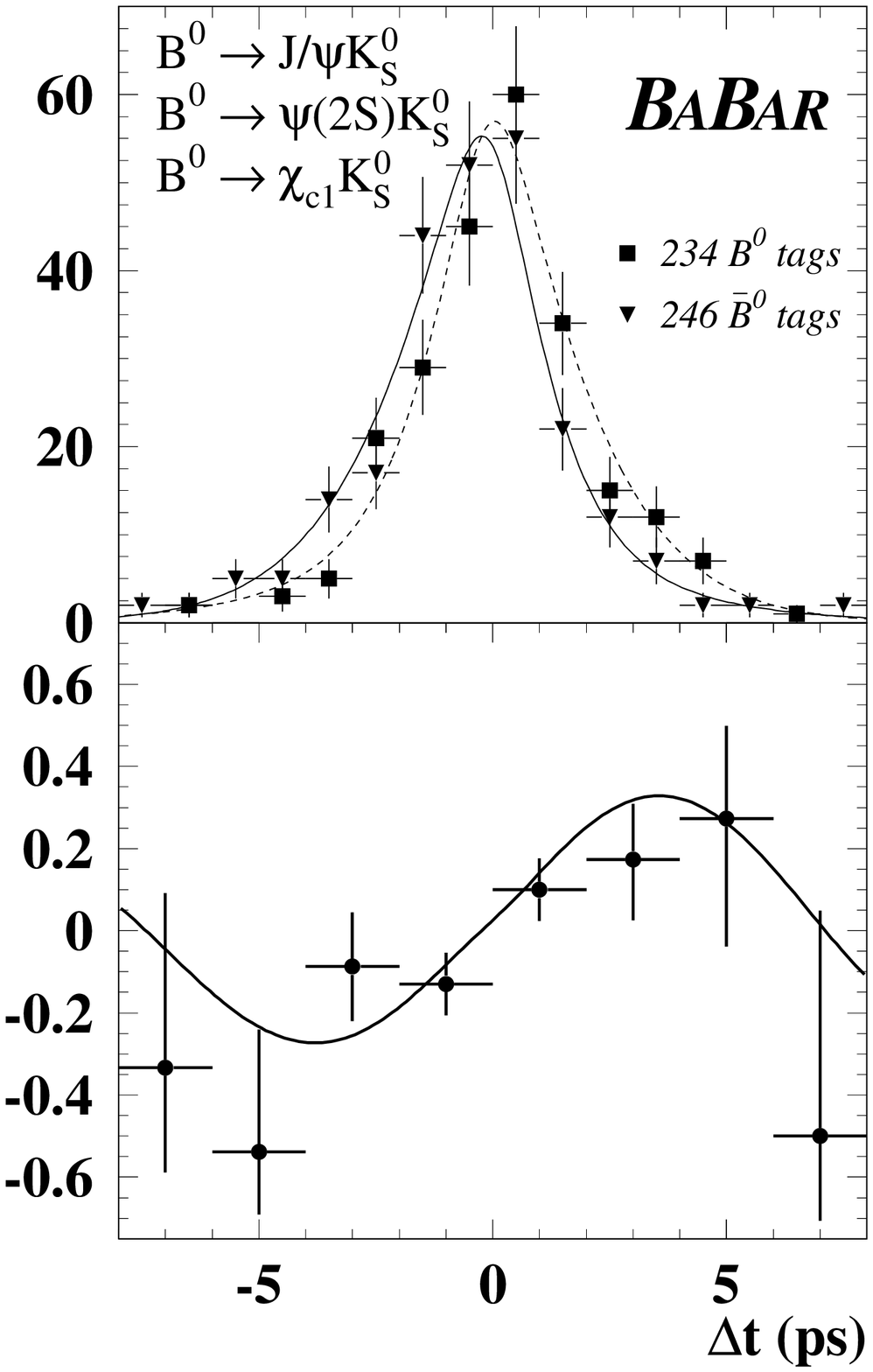, width=10cm} \caption{Time distributions and
raw asymmetry for \babar's $\eta_{\CP}=-1$ sample with,
superimposed, the projection of the un-binned maximum likelihood
fits. The top plot shows the time distributions for the \Bz-tagged
(triangles) and \Bzb-tagged \CP\ samples (squares). The bottom
plot shows the resulting raw asymmetry, which is diluted with
respect to Eq.~\ref{eq:cp_asym} due to unperfect flavor tagging
and finite time resolution. \label{fig:babar_dt}}
\end{center}
\end{figure}

The first results  of \babar\ and \belle\ concerning \B\ lifetimes
and \deltambd\ are not only consistent with world
averages~\cite{ref:pdg}, but in many cases competitive in
precision with the combination of all previous results.  This
constitutes an important demonstration of the feasibility of
time-dependent studies at asymmetric $B$ factories and validates
the experimental technique for the \stwob\ measurement. Important
byproducts of these measurements are the determination of the
parameters of the time resolution function and of the mis-tagging
probabilities.  {\it These quantities are evaluated from the data
on the high-statistics flavor sample rather than determined from
Monte-Carlo simulation.}

\begin{figure}[htb]
\begin{center}
\epsfig{file=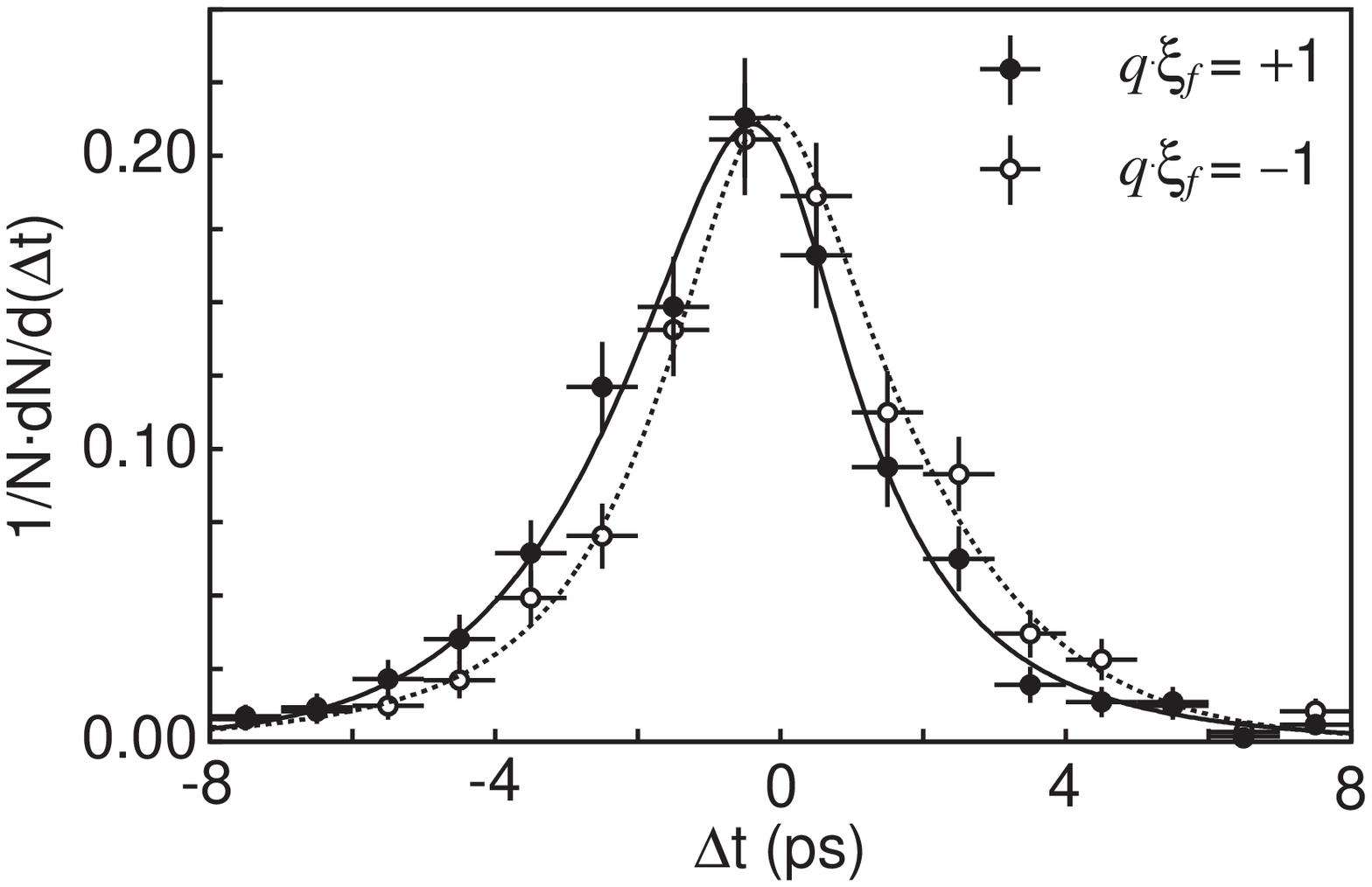, width=12cm} \caption{Time distributions for
\belle's full \CP\ sample after background substraction, with,
superimposed, the projection of the un-binned maximum likelihood
fit to each distribution  (open circles: $\varepsilon_{\rm tag}
\times \eta_{\CP}=-1$ tags; full circles: $\varepsilon_{\rm tag}
\times \eta_{\CP}=+1$ tags). \label{fig:belle_dtdn}}
\end{center}
\end{figure}

The results from \babar~\cite{ref:touramanis}\cite{ref:babar_s2b}
and \belle~\cite{ref:belle_s2b} on \stwob\ are:
\begin{eqnarray}
 \stwob_{\babar} &=& 0.59 \pm 0.14 \pm 0.05  \nonumber \\
 \stwob_{\belle} &=& 0.99 \pm 0.14 \pm 0.06  \nonumber
\end{eqnarray}
These two results independently establish \CP\ violation in the
\B\ meson system at more than 4 standard deviations.  (There is an
unpleasant $ 2 \sigma$ discrepancy between the two results which
will hopefully be resolved with increased statistics.)  The \CP\
violation effect is large, and can be seen from the raw time
distributions (and the resulting time-dependent asymmetry) as a
clear excess of $\varepsilon_{\rm tag} \times \eta_{\CP}=-1$ tags
(resp. $\varepsilon_{\rm tag} \times \eta_{\CP}=+1$ tags) at
positive (resp. negative) $\Delta t$ values (see
Fig.~\ref{fig:babar_dt} and \ref{fig:belle_dtdn}).

Various cross-checks have been performed, including internal
consistency of various sub-samples and tagging categories, null
asymmetry with high statistics charged and $\B_{\rm flav}$
samples, etc. Systematic uncertainties are small since the time
resolution and the tagging performance are extracted from the data
themselves. Main residual systematic uncertainties come from
resolution models, vertex algorithms, detector misalignments and
possible difference between the flavor and the \CP\ samples. The
contribution to the systematics from the fraction, shape and \CP\
content of the background is sizable for the $\jpsi \KL$ and
$\jpsi K^{*0}$ modes, but is negligible overall. Systematics from
the uncertainty on $\tau_{B_d}$ and \deltambd\ are negligible.

\begin{figure}[htb]
\begin{center}
\epsfig{file=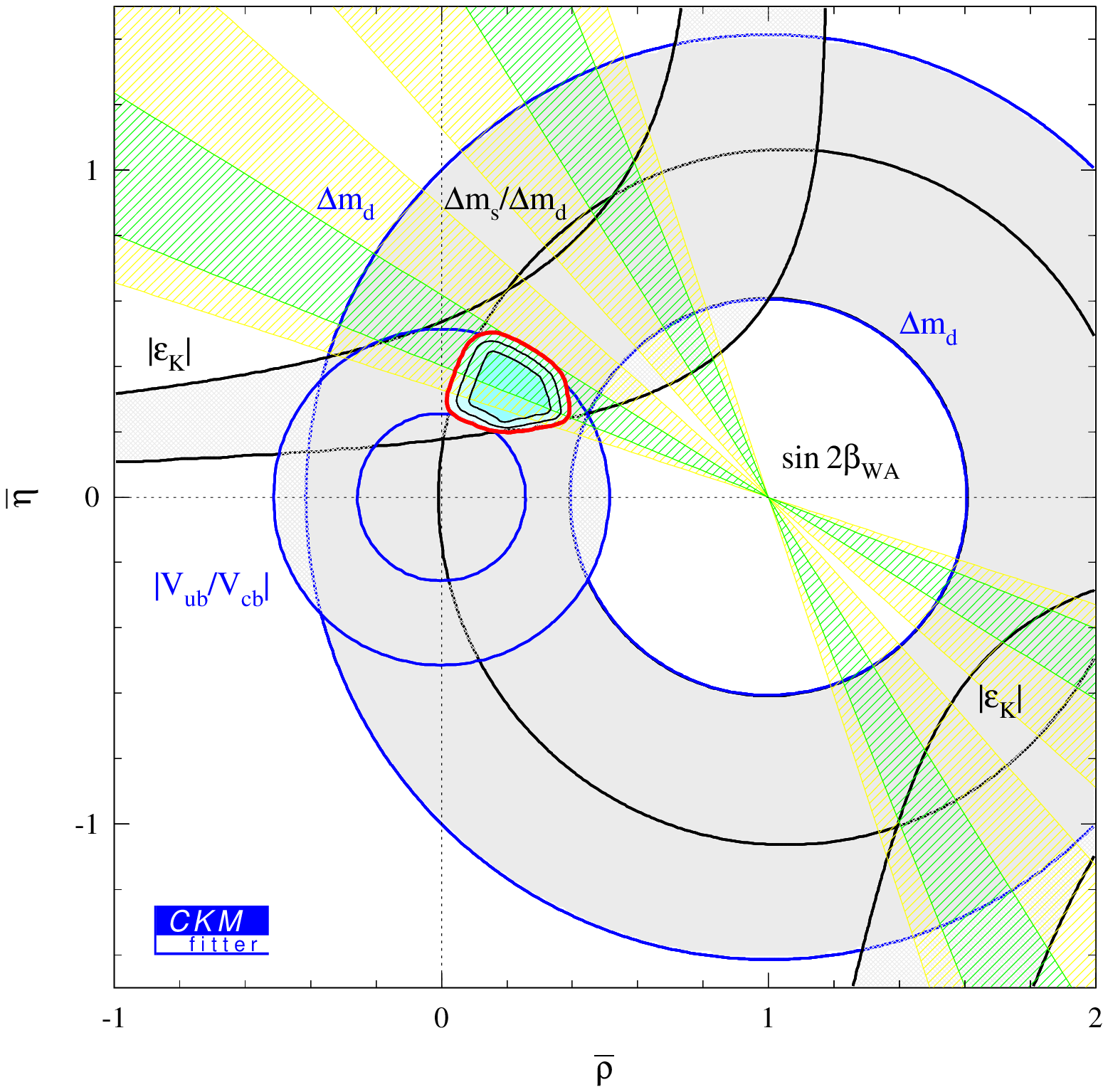, width=12cm} \caption{Present constraints on
the position of the apex of the Unitarity Triangle in the
$(\bar{\rho},\bar{\eta})$ plane~\cite{ref:hoecker}. The world
average of direct measurements of \stwob, represented by green and
yellow hatched regions corresponding to one and two statistical
standard deviations, is not included in the determination of the
allowed region for the apex of the Unitarity Triangle.
\label{fig:rhoeta_wa}}
\end{center}
\end{figure}

Averaging these results with previous measurements from CDF, ALEPH
and OPAL, one obtains  $\stwob_{\rm WA} = 0.79 \pm 0.11 {\rm
(stat+syst)}$.  Figure~\ref{fig:rhoeta_wa} shows how the present
knowledge of the \stwob\ parameter constrains the apex of the
Unitarity Triangle in the $(\bar{\rho},\bar{\eta})$ plane. The
constraint from direct \stwob\ measurements is represented by four
branches, due to the four-fold ambiguity in deriving a value of
$\beta$ from a measurement of \stwob; one of the four possible
solutions, in the upper left quadrant, is obviously consistent
with the allowed region determined from the interpretation of
previous experimental results in the context of the Standard
Model.

\section{Prospects}

At this point, the  knowledge of \stwob\ is not accurate enough to
 reduce the allowed region in the $(\bar{\rho},\bar{\eta})$
plane significantly. However,  \stwob\  constitutes potentially
one of the best constraints on the position of the apex of the
Unitarity Triangle since, as opposed to other quantities, its
measurement is not limited by theoretical uncertainties but by
statistics. Future, highly precise, measurements of \stwob\ will
allow a correspondingly precise test of the CKM model.

The two teams at SLAC and KEK are presently accumulating data at a
unprecedented high rate.  \pepii\ has recently (early October
2001) delivered a record 263\invpb\ of data in one day; \kekb\
holds the record of instantaneous luminosity for an \epem\
collider, $4.9 \times 10^{33} \cm^{-1}\s^{-1}$. If the machines
keep with their schedule, each experiment should have registered
close to $100 \invfb$ of data (with 10\% to 15\% off-resonance) by
early summer 2002. The current results from \belle\ and \babar\
for roughly 30\invfb\ have a statistical uncertainty of
$\sigma_{\stwob}^{\rm stat}\simeq 0.14$. Assuming no further
improvements in reconstruction efficiency and data analysis, we
can infer statistical uncertainties of $\sigma_{\stwob}^{\rm stat}
\simeq 0.08$ for 100\invfb. Most of the systematic uncertainties
will be reduced with larger statistics, and other systematics
dominated by detector effects are likely to improve as well. It is
believed that the total systematic uncertainties can be controlled
at the level of $\sigma_{\stwob}^{\rm syst} \simeq 0.03$ for
100\invfb. The next round of \stwob\ measurements at \B\ factories
will still be statistics dominated.

\subsection{Comparison of \stwob\ in other decay modes}

A comparison of the asymmetry in the pure penguin $b \to
 s\overline{s}s$ decay $B \to \phi \KS$ with that in $b \to
c\overline{c}s$ decays is sensitive to new particles with complex
couplings.  Typically, the experiments can detect one $\phi \KS$
event for 2 \invfb\ of data at the
\FourS~\cite{ref:cleo_phik}\cite{ref:babar_phik}\cite{ref:belle_phik}.
Using the results from the \stwob\ measurement, and scaling from
the observed yields in that mode, the estimated uncertainty using
the $\phi \KS$ mode is $\sigma_{\stwob}^{\rm stat} \simeq 0.56$
for 100\invfb. Clearly, much larger statistics is needed to probe
new physics appearing in loop diagrams by the measurement of
\stwob\ with that mode. Asymmetries in other modes such as $\Bz
\to \jpsi \pi^{0}$ can also potentially measure \stwob, but even
larger statistics are needed.  The $\b \to \c \cbar \d$ modes $\Bz
\to D^{(*)+}D^{(*)-}$ can also (to leading terms) measure \stwob:
most of these modes have been observed by
CLEO~\cite{ref:cleo_dstar} and recently a first measurement of the
polarization in the VV mode $\Bz \to D^{*+}D^{*-}$ has been
presented~\cite{ref:babar_dstar}.

\subsection{Measurements of $\sin{2\alpha}$}

The \CP\ decay mode $\Bz \to \pi^+ \pi^-$ receives competing
contributions from tree and penguin diagrams.   If the decay were
dominated by the tree amplitude, the asymmetry would be
proportional to  $\sin{2 \alpha}$. However, the competing penguin
amplitudes have different weak phases and as a result, the \CP\
asymmetry in this mode measures $ \sin{2\alpha_{\rm eff}} =
\sin{2(\alpha + \delta_{\rm peng})} $, where $\delta_{\rm peng}$
accounts for the penguin contribution. This contribution can in
principle be estimated using isospin relations among $ B \to \pi
\pi $ amplitudes~\cite{ref:gronau-london}. However, the so-called
isospin analysis is in practice jeopardized by the requirement of
measuring flavor-tagged branching ratios $\Bz \to \pi^{0}\pi^{0}$
and $\Bzb \to \pi^{0}\pi^{0}$ (which could be as small as
$10^{-6}$), and by the possible contribution of electroweak
penguin amplitudes. Methods have been developed to bound
experimentally the penguin contribution from a limit on non-tagged
$\Bz \to \pi^0
\pi^0$~\cite{ref:grossman-quinn}-\cite{ref:gronau-london-sinha-sinha}.
Methods allowing the determination of $\delta_{\rm peng}$
theoretically, such as the QCD factorization~\cite{ref:bbns}, are
promising but need to be confronted to
experiment~\cite{ref:beneke_budapest}\cite{ref:ligeti_budapest}.
Experimentally, the measurement is significantly more challenging
than that of \stwob: branching fraction of order $5\times
10^{-6}$, large background from continuum, competition with the
4-times more copious mode $\Bz \to K^+ \pi^-$.  \babar\ has
however performed the first measurement of the time-dependent \CP\
asymmetry in this mode~\cite{ref:babar_s2a}  based on
$65^{+12}_{-11}$ signal $\pi^{+}\pi^{-}$ events selected out of
31~fb$^{-1}$ of on-peak data:
\begin{eqnarray}
\spipi = 0.03^{+0.53}_{-0.56}\pm 0.11, \ \ \cpipi =
-0.25^{+0.45}_{-0.47}\pm 0.14 \, ,
\end{eqnarray}
where \spipi, the coefficient of the sine term (see
Eq.~\ref{eq:asymmetry}), is equal to $ \sin{2\alpha_{\rm eff}}$.
From this result, one can  anticipate with 100\invfb\ an error of
the order of 0.30  on $\sin{2\alpha_{\rm eff}}$  (moreover, it is
conceivable that the $\Bz \to \pi^0 \pi^0$ mode will be observed,
and the $B^+ \to \pi^+ \pi^0$ confirmed).

\subsection{Measurements of $\gamma$}

The study of charmless two-body \B\ decays into pions and kaons
(such as $\B \to K \pi$) offers methods for the determination of
the \CP\ angle $\gamma$.  Non-trivial constraints (bounds) on
$\gamma$ (which is the relative weak phase between tree and
penguin amplitudes for these decays) can be extracted from
\CP-averaged ratios of branching fractions using only isospin
considerations. The general analysis is however complicated by
possible contribution of electroweak penguin amplitudes, SU(3)
breaking effects and final state interactions, so that predictions
suffer from some amount of model dependence (see for instance
Ref.~\cite{ref:bbns} and~\cite{ref:buras-fleischer}). Different
models make predictions on direct \CP\ asymmetries in the $\B \to
K \pi $ modes that can be tested experimentally.

The combination of CKM angles $2\beta + \gamma$ can be measured in
decays of the type $B \to D^{*} \pi$.  Here, the basic idea is to
exploit the interference between the direct decay, $\Bz \to D^{*+}
\pi^-$, and the decay after \BzBzb\ mixing, $\Bz \to \Bzb \to
D^{*+} \pi^-$~\cite{ref:aleksan-dunietz-kayser}.  The difficulty
with this approach is the necessity of measuring the ratio $r$ of
the doubly-Cabibbo-suppressed to the dominant decay amplitudes, in
order to interpret the small expected amplitude of the
time-dependent asymmetry. To overcome the problems specific to
this decay mode, several other strategies based on the
mixing-induced interference between the dominant $\b \to \c \ubar
\d$ and the suppressed $\bbar \to \ubar \c \dbar$ process (the
relative weak phase of which is $\gamma$), have been
proposed~\cite{ref:london-sinha-sinha}\cite{ref:diehl-hiller}.
Experimentally, analyses are based on both full and partial
reconstruction techniques, where the ratio $r$ is determined from
measurements in related less-suppressed modes. Preliminary studies
are encouraging: one can anticipate accuracies
$\sigma_{\sin{(2\beta + \gamma)}}$ of the order of $0.4$ with
100\invfb.

At longer term, the $B^{\pm} \to D K^{\pm}$ modes can be exploited
to extract the \CP\ angle $\gamma$ (the relative weak phase
between the $\b \to \c \ubar \s$ and $\b \to \u \cbar \s$ quark
level amplitudes contributing to these decays). The original
construction suggested in~\cite{ref:gronau-wyler} is using exact
isospin relations which link the modes $B \to \Dz_{\CP} K$, $B \to
\Dz K$ and $B \to \Dzb K$ ($\Dz_{\CP}$ represents a \CP\
eigenstate), but suffers from intrinsic
difficulties~\cite{ref:atwood-dunietz-soni-97}; several variants
of the original method have been
proposed~\cite{ref:atwood-dunietz-soni-01}\cite{ref:gronau-rosner},
but all require the measurement of very rare processes, at the
$10^{-7}$ level. In addition, this type of construction suffers
from an eight-fold ambiguity in the value of $ \gamma$, so even
higher precision is necessary to separate the multiple solutions
for $\gamma$. Experimentally, evidence for $B \to D_{\CP} K$
decays has been reported~\cite{ref:belle_dcpk}.

\section{Conclusions}

The last two years have been rich in important experimental
results on \CP\ violation. The asymmetric \B\ factories have come
online with outstanding beginnings: the \babar\ and \belle\
collaborations have already published world class results in the
domain of \B\ physics, and have established independently \CP\
violation in the \B\ system. Their measurements of \stwob,
obtained with $30\invfb$ each, dominate the present world average,
$\stwob_{\rm WA}= 0.79 \pm 0.11 {\rm (stat+syst)}$. The
encouraging \babar\ result on $\sin{ 2 \alpha_{\rm eff}}$
demonstrates the feasibility of time-dependent \CP\ analyses with
small signals and large backgrounds at \B\ factories.  The data
samples are expected to be multiplied by three before next summer,
and by ten in the next three years, bringing sensitivities to
direct and indirect \CP\ violation effects to the level of
theoretical expectations. Experiments at hadron colliders benefit
from much higher statistics in \B\ mesons (but with different
levels of backgrounds and different systematics), and the
possibility of studying the $B_s$ system.   The upgraded CDF and
D0 experiments at the Tevatron are starting up again and will
produce competitive measurements of \stwob\ and probably the
measurement of $x_s$, as soon as in 2002. The collider experiments
of the next generation, BTeV at the Tevatron and LHCB at the LHC,
are planned to come online around year 2006 .

Complementary approaches of \B\ factories and collider experiments
will be needed to perform the rich programme of redundant
precision measurements in the domain of charged weak interactions
for testing the CKM sector of the Standard Model, and probing the
origin of the \CP\ violation phenomenon.

\section{Acknowledgements}

It is my pleasure to thank Masa Yamauchi, Karl Ecklund, Zoltan
Ligeti, Martin Beneke for their help, as well as  my \babar\
colleagues for their support, in particular Georges London and
Andreas H\"{o}cker.

\end{document}